\documentclass[amsmath,amssymb,reprint,aps,pra,floatfix]{revtex4-1}
\usepackage{bm}
\usepackage{amsmath}
\usepackage{graphicx}
\usepackage{bm}
\usepackage{braket}
\usepackage{epstopdf}
\usepackage{bm}
\usepackage{bbm}
\usepackage{lineno}

\makeatother

\begin{document}

\title{Determining complementary properties using weak-measurement: uncertainty, predictability, and disturbance}

\author{G. S. Thekkadath}
\email{guillaume.thekkadath@physics.ox.ac.uk}
\affiliation{Clarendon Laboratory, University of Oxford, Parks Road, Oxford, OX1 3PU, UK}

\author{F. Hufnagel}
\affiliation{Department of Physics and Centre for Research in Photonics, University
of Ottawa, 25 Templeton Street, Ottawa, Ontario, K1N 6N5, Canada}

\author{J. S. Lundeen}
\email{jlundeen@uottawa.ca}
\affiliation{Department of Physics and Centre for Research in Photonics, University
of Ottawa, 25 Templeton Street, Ottawa, Ontario, K1N 6N5, Canada}

\begin{abstract}
It is often said that measuring a system's position must disturb the complementary property, momentum, by some minimum amount due to the Heisenberg uncertainty principle. Using a ``weak-measurement", this disturbance can be reduced. One might expect this comes at the cost of also reducing the measurement's precision. However, it was recently demonstrated that a sequence consisting of a weak position measurement followed by a regular momentum measurement can probe a quantum system at a single point, with zero width, in position-momentum space. Here, we study this ``joint weak-measurement" and reconcile its compatibility with the uncertainty principle. While a single trial probes the system with a resolution that can saturate Heisenberg's limit, we show that averaging over many trials can be used to surpass this limit. The weak-measurement does not trade-away precision, but rather another type of uncertainty called ``predictability" which quantifies the certainty of retrodicting the measurement's outcome.
\end{abstract}
\maketitle


\section{Introduction}
The Heisenberg uncertainty principle (HUP) plays a central role in
the description of both states and measurements in quantum physics.
In the former, the HUP refers to an intrinsic limit in the precision
with which a system can be prepared to simultaneously have some position
$x$ and momentum $p$~\cite{kennard1927quantenmechanik,robertson1929uncertainty}. Even with independent
measurements of these properties on identical and separate copies of the
system, one would always find a spread in their measurement statistics
that satisfies $\Delta x\Delta p\geq 1/2$ (we use $\hbar \equiv 1$ throughout the paper). As a result, quantum
states cannot be represented in phase space (i.e. $x$-$p$ space) by
a single point. Instead, they are described by quasiprobability distributions $W(x,p)$
such as the Wigner function. These have non-classical features (e.g.
negative probabilities) that prevent complementary properties like $x$ and $p$ from
being simultaneously specified with an arbitrary precision. This understanding
of the HUP is uncontroversial and is taught in undergraduate physics
courses~\cite{sakurai2017modern}.

In contrast, the significance of the HUP in measurements on a single copy of a system is a contentious topic~\cite{Scully1991quantum,Storey1994path,Wiseman1995uncertainty}. Heisenberg originally derived the HUP by considering the momentum kick $\epsilon_p$ imparted onto an electron by a position measurement of precision $\eta_x$~\cite{heisenberg1927quantum}. He found that $\eta_x\epsilon_p\geq 1/2$. This thought-experiment, called Heisenberg's microscope, provides an intuitive understanding of the HUP: there is a trade-off between the disturbance $\epsilon_p$ and precision $\eta_x$ of the position measurement. While this intuition is correct, one can derive tighter ``error-disturbance" bounds on $\eta_x\epsilon_p$ than the HUP~\cite{ozawa2004uncertainty,ozawa2003universally,branciard2013error,busch2013proof,busch2014colloquium}.

There are several shortcomings with such error-disturbance bounds. Firstly, there is no consensus on how disturbance $\epsilon_p$ should be defined~\cite{rozema2015note}. Secondly, $\epsilon_p$ and $\eta_x$ are usually determined by averaging the measurement over many trials (see Refs.~\cite{hofmann2003uncertainty, dressel2014certainty} for counter-examples). As such, error-disturbance relations do not provide much insight as to how precisely one can \textit{simultaneously} (i.e. jointly) measure $x$ and $p$ in a single trial. 

An ideal joint measurement of position and momentum determines whether a system is at a particular $x$ and $p$, i.e. the joint (quasi)probability $W(x=x', p=p')$. By repeating this joint measurement while scanning $x'$ and $p'$, one could in principle fully determine the state $W(x,p)$ of a general system. Techniques to perform such a joint measurement have been been continuously investigated since the inception of quantum physics~\cite{Pauli1980,arthurs1965simultaneous,she1966simultaneous,Park1968simultaneous,muynck1979simultaneous,arthurs1988quantum,leonhardt1993phase,hall2004prior,hofmann2012how,hofmann2014sequential,thekkadath2017determining}. Naively, one might think the joint measurement could be achieved by simultaneously measuring the projectors $\bm{\pi}_{x'} = \ket{x'}\bra{x'}$ and $\bm{\pi}_{p'} = \ket{p'}\bra{p'}$. However, these projectors do not commute. If one were to try to measure them sequentially, e.g. by first measuring $\bm{\pi}_{x'}$, the position measurement would disturb the system's momentum. But what if the system's position is ``weakly'' measured as to not disturb its momentum? That is, consider a sequence consisting of a weak-measurement of $\bm{\pi}_{x'}$ followed by a regular measurement of $\bm{\pi}_{p'}$. Henceforth, we refer to this sequence as a joint weak-measurement (JWM). 

Using the same intuition as in Heisenberg's microscope, one would expect that the weak-measurement of $\bm{\pi}_{x'}$ must trade away its precision, e.g. turn into a measurement with a finite width in position. However, rather counter-intuitively, it was recently shown that the average outcome of a JWM determines the value of the system's state in phase space at a single point $(x,p)=(x',p')$~\cite{lundeen2012procedure}. Indeed, it has
been experimentally demonstrated that the average outcome of a JWM directly gives the
wavefunction~\cite{Lundeen2011direct} or quasiprobability distribution~\cite{salvail2013full,bamber2014observing} of the measured state. Beyond their application in state determination, JWMs have been used to probe foundational issues in quantum physics~\cite{lundeen2009experimental,yokota2009direct,kocsis2011observing,rozema2012violation,ringbauer2014joint,suzuki2016observation,mahler2016experimental,curic2018experimental}. 

The fact that a JWM can probe a single point in phase space conflicts with the intuitive arguments above: it suggests that the weak-measurement is not trading away its precision. Here, we shed some light on this issue. The paper is structured as follows. In Sec.~\ref{sec:jwm}, we derive a measurement operator describing the JWM. We show that the JWM projects the system onto a coherent superposition of a position and momentum eigenstate with a relative weight given by a quantity called the ``predictability". In Sec.~\ref{sec:psd}, we study the JWM in phase space. We derive the Wigner function of the JWM operator and discuss its features. We also comment on the compatibility of the JWM with the HUP. In Sec.~\ref{sec:unc_pred}, we discuss the physical meaning of the predictability.


\section{Joint weak-measurement}
\label{sec:jwm}
The joint weak-measurement (JWM) consists of a sequence of two measurements on a system $\mathcal{S}$ prepared in a state $\psi^{\mathcal{S}}(x) = \braket{x^{\mathcal{S}}|\psi^{\mathcal{S}}}$. The first is a weak-measurement of position,
$\bm{\pi}_{x'}^{\mathcal{S}} = \ket{x'^{\mathcal{S}}}\bra{x'^{\mathcal{S}}}$ (we use this notation for projectors throughout the paper), followed
by a regular measurement of momentum, $\bm{\pi}_{p'}^{\mathcal{S}}$.

The concept of weak-measurement was introduced in Refs.~\cite{aav1988how,aharonov1990properties}. The weak position measurement is implemented by weakly coupling the observable $\bm{\pi}_{x'}^{\mathcal{S}}$
in $\mathcal{S}$ to a ``pointer'' observable $\bm{\alpha}^{\mathcal{A}}$
in an ancillary system $\mathcal{A}$ (the ``ancilla'') through:
\begin{equation}
\bm{U}_{x'}=\exp{(-i\gamma\bm{\pi}_{x'}^{\mathcal{S}}\otimes\bm{\alpha}^{\mathcal{A}})},\label{eqn:vonNeumann}
\end{equation}
where $\gamma$ is the strength of the interaction (since $\hbar\equiv1$, $\gamma$ has units of
length). For the sake of
definiteness, we choose $\mathcal{A}$ to be a separate
particle that has a Gaussian wavefunction with position $q$,
width $\sigma$, and is initially centered at $q=-\gamma /2$, i.e. $\phi^{\mathcal{A}}(q+\gamma/2)=e^{-(q+\gamma/2)^{2}/2\sigma^{2}}/(\pi\sigma^{2})^{1/4}$.
We choose $\bm{\alpha}^{\mathcal{A}}$ to be the particle's momentum
operator, $\bm{\alpha}^{\mathcal{A}}=-i\partial/\partial q$. Thus, the action of $\bm{U}_{x'}$ is to shift the center position of
the ancilla wavefunction by an amount proportional to the outcome
of the $\bm{\pi}_{x'}^{\mathcal{S}}$ measurement, i.e. $\phi^{\mathcal{A}}(q+\gamma/2)\rightarrow\phi^{\mathcal{A}}(q\pm\gamma/2)$,
with $+$ and $-$ respectively corresponding to the $\bm{\pi}_{x'}^{\mathcal{S}}$
eigenvalues $0$ ($x\neq x'$) and $1$ ($x=x'$). For later reference,
the respective probability distributions are:
\begin{equation}
\begin{split}
P(q|x=x')&=\left|\phi^{\mathcal{A}}(q-\gamma/2)\right|^{2} \\
P(q|x\neq x')&=\left|\phi^{\mathcal{A}}(q+\gamma/2)\right|^{2}.
\label{eq:prob_pointer}
\end{split}
\end{equation}

The interaction $\bm{U}_{x'}$ entangles the ancilla with the system
thereby allowing measurements on the ancilla to be correlated with
the state of the system. In the strong (i.e. regular) measurement
limit ($\gamma\gg\sigma$), the ancilla's shift unambiguously indicates
the outcome of the $\bm{\pi}_{x'}^{\mathcal{S}}$ measurement. However,
because of the entanglement, measuring the ancilla's position also
disturbs the state of the system. In particular, it destroys the coherence
between the amplitude for the $x=x'$ position with the amplitude for the
remaining $x\neq x'$ positions. This disrupts subsequent measurements (e.g. of momentum).
In contrast, in the weak measurement limit ($\gamma\ll\sigma$), the
ancilla's shift lies within its initial position distribution
and thus cannot be resolved in a single trial. The benefit is that,
in each trial, the entanglement, and thus disturbance, is minimized.
Consequently, subsequent measurements can reveal faithful information
about the system's initial state $\psi^{\mathcal{S}}(x)$. In both
the strong and weak measurement limits, the average result of the
measurement of $\bm{\pi}_{x'}^{\mathcal{S}}$ over many trials can be found by determining the average ancilla shift, i.e. $\braket{\bm{q}^\mathcal{A}} = \gamma \braket{\bm{\pi}_{x'}^\mathcal{S}}$.

In each trial, subsequent to the weak-measurement interaction $\bm{U}_{x'}$,
we also perform a measurement of the system's momentum, $\bm{\pi}^{\mathcal{S}}_{p'}$. Since there are no subsequent measurements to disrupt, this last momentum measurement can be strong. In that case, the joint probability of measuring the ancilla to have position $q=q'$
and system to have momentum $p=p'$ is: 
\begin{equation}
P_{x'}(q',p')=\bra{\phi^{\mathcal{A}}}\bra{\psi^{\mathcal{S}}}\bm{U}_{x'}^{\dagger}\bm{\pi}_{q'}^{\mathcal{A}}\bm{\pi}_{p'}^{\mathcal{S}}\bm{U}_{x'}\ket{\psi^{\mathcal{S}}}\ket{\phi^{\mathcal{A}}}.
\label{eqn:prob_weak}
\end{equation}

So far, we have described the JWM in terms of projective measurements in
the system-ancilla Hilbert space $\mathcal{S}\otimes\mathcal{A}$.
To describe the action of the JWM on the system alone, we can
define a measurement operator $\bm{M}_{q',x',p'}^{\mathcal{S}}$ which acts solely in $\mathcal{S}$
but fully reproduces the statistics $P_{x'}(q',p')$~\cite{wiseman2010quantum}: 
\begin{equation}
P_{x'}(q',p')=\bra{\psi^{\mathcal{S}}}\bm{M}_{q',x',p'}^{\mathcal{S}}\ket{\psi^{\mathcal{S}}}.
\label{eqn:povm_defn}
\end{equation}
By comparing Eq.~\eqref{eqn:povm_defn} with Eq.~\eqref{eqn:prob_weak},
it is clear that $\bm{M}_{q',x',p'}^{\mathcal{S}}=\bra{\phi^{\mathcal{A}}}\bm{U}_{x'}^{\dagger}\bm{\pi}_{q'}^{\mathcal{A}}\bm{\pi}_{p'}^{\mathcal{S}}\bm{U}_{x'}\ket{\phi^{\mathcal{A}}}$.
In appendix~\ref{app:A}, we expand this equation and show that  $\bm{M}_{q',x',p'}^{\mathcal{S}} = |\phi^{\mathcal{A}}(q')|^2\ket{x'^{\mathcal{S}},p'^{\mathcal{S}}}_{q'}\bra{x'^{\mathcal{S}},p'^{\mathcal{S}}}_{q'}$, i.e. a projector onto the state $\ket{x'^{\mathcal{S}},p'^{\mathcal{S}}}_{q'}$ weighted by the probability for the ancilla to be at position $q'$.  In the weak measurement limit $\gamma/\sigma \ll 1$, the state $\ket{x'^{\mathcal{S}},p'^{\mathcal{S}}}_{q'}$ is:
\begin{equation}
\ket{x'^{\mathcal{S}},p'^{\mathcal{S}}}_{q'}\approx \ket{p'^{\mathcal{S}}}+\mathcal{P}_{q'}e^{ip'x'}\ket{x'^{\mathcal{S}}},
\label{eqn:approx_proj}
\end{equation}
where $|\mathcal{P}_{q'}|\equiv\gamma |q'|/\sigma^{2}$ is a factor
called the ``predictability.'' We discuss the physical meaning
of this factor later. The state in Eq.~\eqref{eqn:approx_proj} is rather unusual: it is a coherent superposition of a position eigenstate $\ket{x'^{\mathcal{S}}}$ and a momentum eigenstate $\ket{p'^{\mathcal{S}}}$. Because $\bm{\pi}^{\mathcal{S}}_{x'}$ is weakly measured, the eigenstate $\ket{x'^{\mathcal{S}}}$ in Eq.~\eqref{eqn:approx_proj} is weighted by a factor containing the weak-measurement strength $\gamma/\sigma$, as might be expected. On the other hand, the state in Eq.~\eqref{eqn:approx_proj} is truly unusual since it contains coherence \textit{between} position and momentum. Typically, coherence is considered within one or the other of these spaces, not between them. Such states have not received much attention in the literature, which is not surprising given that it was hitherto unclear how to prepare them or project onto them. One exception is Ref.~\cite{hofmann2017quantum} which shows that a particle in a state like Eq.~\eqref{eqn:approx_proj} can violate Newton's first law. The coherence between the particle's position and momentum allows for interference between the two properties. As a result, the particle is not restricted to move along a straight trajectory. 

We note that the unusual form of Eq.~\eqref{eqn:approx_proj} is not an artifact of the fact that we derived the JWM operator using projectors onto single eigenstates, i.e. $\bm{\pi}^{\mathcal{S}}_{x'} =\ket{x'^{\mathcal{S}}}\bra{x'^{\mathcal{S}}} $ and $\bm{\pi}^{\mathcal{S}}_{p'} = \ket{p'^{\mathcal{S}}}\bra{p'^{\mathcal{S}}}$. We can generalize the JWM by considering measurements that project the system onto finite-width position and momentum distributions, i.e. $\bm{\pi}^{\mathcal{S}}_{x'} \rightarrow \bm{\Pi}^{\mathcal{S}}_{x'} = \ket{\chi^{\mathcal{S}}(x')}\bra{\chi^{\mathcal{S}}(x')}$ and $\bm{\pi}^{\mathcal{S}}_{p'} \rightarrow \bm{\Pi}^{\mathcal{S}}_{p'} = \ket{\Gamma^{\mathcal{S}}(p')}\bra{\Gamma^{\mathcal{S}}(p')}$. In appendix~\ref{app:B}, we show that taking such projectors into account has the effect of transforming the single eigenstates in Eq.~\eqref{eqn:approx_proj} into the corresponding finite-width distributions. That is, in the weak limit, this more general JWM projects the system onto the state:
\begin{equation}
\ket{\Delta^{\mathcal{S}}_{q'}(x',p')} \approx \ket{\Gamma^{\mathcal{S}}(p')} + \mathcal{P}_{q'} \braket{\chi^{\mathcal{S}}(x')|\Gamma^{\mathcal{S}}(p')} \ket{\chi^{\mathcal{S}}(x')},
\label{eqn:approx_proj_general}
\end{equation}
which is a generalization of the state in Eq.~\eqref{eqn:approx_proj}. Both states share the unusual features just discussed such as coherence between position and momentum. In the next section, we study the Wigner function of the state in Eq.~\eqref{eqn:approx_proj_general} to clarify its physical significance and visualize its unusual features in phase space.

\section{Phase space description}
\label{sec:psd}
\subsection{Motivation}
In general, both quantum states (i.e. density matrices $\bm{\rho}$) and measurements (i.e. elements $\bm{M}$ of a positive-operator valued measure) can be described by a positive semi-definite Hermitian operator $\bm{O}$~\cite{wiseman2010quantum}. This duality between states and measurements is even clearer in a phase space description. For instance, the Wigner function of $\bm{O}$ is found through the inverse Weyl transformation: $W_{\bm{O}}(x,p)= \int_{-\infty}^{\infty}dy\braket{x+y|\bm{O}|x-y}e^{-i2py} / \pi$~\cite{wigner1932on,case2008wigner}. In this framework, the average outcome of a measurement is determined by the overlap between the measurement and state Wigner functions, e.g. $\braket{\bm{M}} = \mathrm{Tr}(\bm{M}\bm{\rho}) = \int_{-\infty}^{\infty}\int_{-\infty}^{\infty}dxdp W_{\bm{M}}(x,p)W_{\bm{\rho}}(x,p)$. This makes Wigner functions a useful tool to visualize the action of measurements~\cite{lundeen2009tomography}. 

Moreover, the Wigner function provides a straightforward way to understand the significance of the HUP in both states and measurements. In both cases, the variances in the marginals of a Wigner function must satisfy the HUP, $\Delta^{2}x\Delta^{2}p\geq1/4$~\cite{narcowich1986necessary}. For example, the Wigner function of a coherent state $\bm{\rho} = \ket{\alpha}\bra{\alpha}$ with complex amplitude $\alpha = \mathrm{Re}[\alpha] + i\mathrm{Im}[\alpha]$, i.e. $W_{\bm{\rho}}(x,p)=\exp{\left(-(x-\mathrm{Re}[\alpha])^{2}-(p-\mathrm{Im}[\alpha])^{2}\right)}/\pi$, saturates the HUP: $\Delta^{2}x\Delta^{2}p=1/4$. Here, the marginal variances $\Delta^{2}x$ and $\Delta^{2}p$ express the spread in the system's position and momentum, respectively. The measurement-equivalent to the coherent state, i.e. $\bm{M}=\ket{\alpha}\bra{\alpha}$, can be achieved using eight-port homodyne detection~\cite{leonhardt1993phase}. The marginal variances of the measurement Wigner function $W_{\bm{M}}(x,p)$ express how precisely $x$ and $p$ are simultaneously probed by $\bm{M}$.

\subsection{Wigner function of the joint weak-measurement}

\begin{figure}
\centering 
\includegraphics[width=0.5\textwidth]{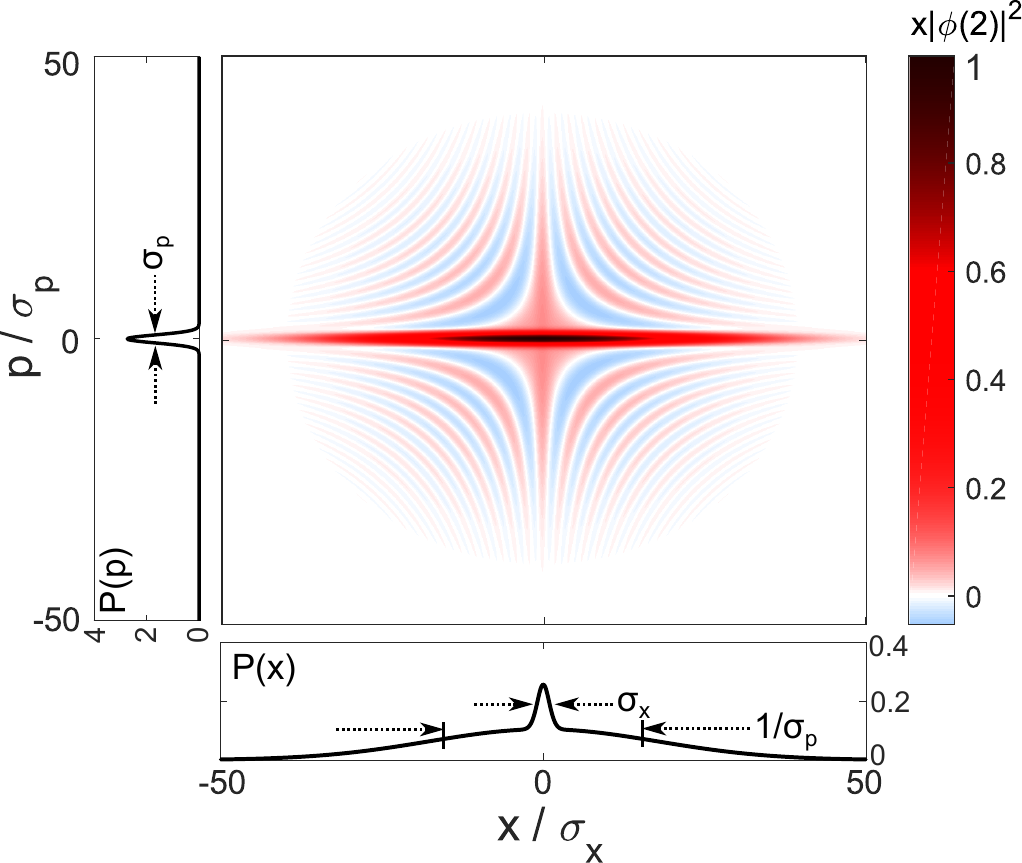} 
\caption{The Wigner function and marginals of the joint weak-measurement. Here we consider $\sigma_x = \sigma_p = 0.2$, $x'=0$, $p'=0$, $q'=2$, $\gamma=0.2$, and $\sigma=1$, giving a predictability of $\mathcal{P}_{q'=2}=0.4$, which occurs in at most $|\phi^{\mathcal{A}}(2)|^{2}\approx1\%$ of trials. We plot $W_{\bm{M}}(x,p)$ (given in Eq.~\eqref{eqn:general_wig}) scaled by $|\phi^\mathcal{A}(2)|^2$. We also plot the marginal distributions $P(x)$ and $P(p)$ in black (given in Eq.~\eqref{eqn:general_marginals}). The broad background of width $\sim 1/\sigma_p$ in $P(x)$ is eliminated when the JWM is averaged over many trials.}
\label{fig:3dWigner} 
\end{figure}

The previous paragraph motivates studying the JWM in phase space. We focus on the general JWM, i.e. Eq.~\eqref{eqn:approx_proj_general}. The corresponding measurement operator is $\bm{M}_{q',x',p'}^{\mathcal{S}} = |\phi^{\mathcal{A}}(q')|^2 \ket{\Delta^{\mathcal{S}}_{q'}(x',p')} \bra{\Delta^{\mathcal{S}}_{q'}(x',p')}$. The Wigner function of the JWM operator $W_{\bm{M}}(x,p)$ is derived in appendix~\ref{app:C} assuming that $\chi^{\mathcal{S}}(x)$ and $\Gamma^{\mathcal{S}}(p)$ are Gaussians with respective widths $\sigma_x$ and $\sigma_p$ that are narrow, $\sigma_x\sigma_p \ll 1$. The result is plotted in Fig.~\ref{fig:3dWigner} for the particular case of $x'=0$ and $p'=0$. In the general case, $W_{\bm{M}}(x,p)$ resembles a cross centered at the measurement probe location, $(x,p)=(x',p')$. The cross is composed of two squeezed coherent states, i.e. Gaussians with $\Delta x \Delta p = 1/2$ but $\Delta x \neq \Delta p$, each describing one of the two projective measurements in the JWM sequence. The first, with $\Delta p \gg \Delta x$, corresponds to the weak measurement of $\bm{\Pi}^{\mathcal{S}}_{x'} $. It is centered along
the line $x=x'$ and is scaled by $\mathcal{P}_{q'}^{2} = 0.16$. The second, with $\Delta x \gg \Delta p$,
corresponds to a regular measurement of $\bm{\Pi}^{\mathcal{S}}_{p'} $. It is centered along the line $p=p'$. 

However, the joint measurement cannot be simply be explained as an
incoherent combination of these two projectors.
Such a measurement would correspond to independently, rather than
jointly, measuring the projectors $\bm{\Pi}^{\mathcal{S}}_{x'}$ and $\bm{\Pi}^{\mathcal{S}}_{p'}$. Instead, due to the coherence between the two projectors in the JWM sequence, there is an interference term leading to negativity and fringes in $W_{\bm{M}}(x,p)$. Negativity is considered to be a sign of non-classicality~\cite{kenfack2004negativity,spekkens2008negativity} and a resource for quantum information processing~\cite{Howard2014contextuality}.

\subsection{Single trial marginal variances}

The marginals of $W_{\bm{M}}(x,p)$ are given by $P(x) = \left|\braket{x|\Delta^{\mathcal{S}}_{q'}(x',p')}\right|^2$ and $P(p) = \left|\braket{p|\Delta^{\mathcal{S}}_{q'}(x',p')}\right|^2$ (we normalized the two by dividing them by $|\phi^{\mathcal{A}}(q')|^2$). These are plotted in Fig.~\ref{fig:3dWigner}. In appendix~\ref{app:D}, we show that, for $\sigma_x\sigma_p \ll 1$, the respective variances of $P(x)$ and $P(p)$ are:
\begin{equation}
\begin{aligned}
\Delta^2 x &=  1/(2\sigma^2_p) + \mathcal{P}_{q'}^2\sigma_x^3\sigma_p + 4\mathcal{P}_{q'}\sigma_x^3\sigma_p \\
\Delta^2 p &= \sigma^2_p/2 + \mathcal{P}_{q'}^2\sigma_p/\sigma_x + 4\mathcal{P}_{q'}\sigma_p^3\sigma_x .
\end{aligned}
\label{eqn:marginals1}
\end{equation}
These marginal variances quantify the precision of a single trial of the JWM in which the ancilla is measured to have position $q'$. One can check that $\Delta^2 x\Delta^2 p \geq 1/4$ when $\sigma_x\sigma_p \ll 1$, and hence the HUP is satisfied. In the limit $\sigma_p, \sigma_x \rightarrow 0$, the JWM projects the system onto exact position and momentum eigenstates (as in Eq.~\eqref{eqn:approx_proj}), in which case $\Delta^2 x \rightarrow \infty$ and $\Delta^2 p \rightarrow 0$. Moreover, in the limit $|\mathcal{P}_{q'}| \rightarrow 0$, i.e. when the predictability vanishes, the JWM saturates the HUP regardless of $\sigma_p$. The physical significance of this limit is discussed in Sec.~\ref{sec:unc_pred}.

\subsection{Averaging}
Since the coupling between the ancilla and system is weak, the outcome of the weak-measurement is ambiguous, i.e. $P(q|x=x')$ and $P(q|x\neq x')$ are overlapping. As such, in a single measurement trial, the measured ancilla position $q'$ does not determine with certainty whether $x=x'$ or $x\neq x'$. To overcome this, one can repeat many trials in order to find the average ancilla position shift $\braket{\bm{q}^{\mathcal{A}}}$. This averaging is the standard procedure in weak measurement. The quantity $\braket{\bm{q}^{\mathcal{A}}}$ unambiguously determines the result of the JWM:
\begin{equation}
\braket{\bm{q}^{\mathcal{A}}}=\int_{-\infty}^{\infty}q'\braket{\bm{M}_{q,x',p'}^{\mathcal{S}}}dq'=\gamma\mathrm{Re}\braket{\bm{\Pi}_{p'}^{\mathcal{S}}\bm{\Pi}_{x'}^{\mathcal{S}}}.
\label{eqn:ancilla_dirac}
\end{equation}
Recall that for $\sigma_p, \sigma_x \rightarrow 0$, the JWM project onto single eigenstates, i.e. $\bm{\Pi}_{x'}^{\mathcal{S}}\rightarrow \bm{\pi}_{x'}^{\mathcal{S}}$ and $\bm{\Pi}_{p'}^{\mathcal{S}}\rightarrow \bm{\pi}_{p'}^{\mathcal{S}}$. In this limit, we find that $\braket{\bm{q}^{\mathcal{A}}} = \gamma \mathrm{Re}\braket{\bm{\pi}_{p'}^{\mathcal{S}}\bm{\pi}_{x'}^{\mathcal{S}}}$ using Eq.~\eqref{eqn:ancilla_dirac}. Expanding this last quantity, one finds $\braket{\bm{\pi}_{p'}^{\mathcal{S}}\bm{\pi}_{x'}^{\mathcal{S}}} = \psi^{\mathcal{S}}(x')\tilde{\psi}^{*\mathcal{S}}(p')e^{-ip'x'}\equiv D(x',p')$ where $\tilde{\psi}^{\mathcal{S}}(p')$ is the Fourier transform of $\psi^{\mathcal{S}}(x')$ and $^*$ denotes the complex conjugate. The quantity $D(x',p')$ is a quasiprobability distribution of the state $\psi^{\mathcal{S}}(x)$ called the Dirac distribution~\cite{kirkwood1933quantum,dirac1945on}. Much like the Wigner function, the Dirac distribution fully describes the state $\psi^{\mathcal{S}}(x')$ in phase space. Since $\braket{\bm{q}^{\mathcal{A}}} = \gamma \mathrm{Re}[D(x',p')]$, the average outcome of the JWM probes phase space at a single point. We note that $\mathrm{Im}[D(x',p')]$ can be obtained by instead determining the average momentum shift of the ancilla~\cite{lundeen2012procedure}.

As opposed to $W_{\bm{M}}(x,p)$, the average JWM Wigner function $W_{\braket{\bm{q}}}(x,p) = \int_{-\infty}^{\infty}dq' q' W_{\bm{M}}(x,p)$ no longer looks like a cross in phase space. The averaging procedure eliminates the two squeezed terms in $W_{\bm{M}}(x,p)$. Thus, $W_{\braket{\bm{q}}}(x,p)$ consists only of the coherence between the two projectors in the JWM sequence, i.e. the term containing the fringes and negativity in phase space.

Since the averaged JWM can probe phase space at a single point, we expect that the marginal variances of $W_{\braket{\bm{q}}}(x,p)$ should vanish. We obtain the variances of this averaged JWM by averaging the variances in Eq.~\eqref{eqn:marginals1} over all possible ancilla positions $q'$ weighted by their corresponding probability $|\phi^{\mathcal{A}}(q')|^2$ (and normalizing by the interaction strength $\gamma$). That is,
\begin{equation}
\begin{aligned}
\Delta^2 x_{\braket{\bm{q}}} & =\int_{-\infty}^{\infty}dq'(\Delta^2 x)q'|\phi^{\mathcal{A}}(q')|^2/\gamma = 2\sigma_x^3\sigma_p, \\
\Delta^2 p_{\braket{\bm{q}}} & =\int_{-\infty}^{\infty}dq'(\Delta^2 p)q'|\phi^{\mathcal{A}}(q')|^2/\gamma = 2\sigma_p^3\sigma_x, \\
\end{aligned}
\label{eqn:marginals2}
\end{equation}
which leads to $\Delta^2 x_{\braket{\bm{q}}}\Delta^2 p_{\braket{\bm{q}}} = 4\sigma_x^4\sigma_p^4$. In the limit $\sigma_p, \sigma_x \rightarrow 0$, this product vanishes. Comparing this result with Eq.~\eqref{eqn:marginals1} in the same limit, we see that averaging enables the JWM to probe the position and momentum of a system with a precision exceeding the HUP. Indeed, looking at Fig.~\ref{fig:3dWigner}, there is a broad background in $P(x)$ that has width $\sim 1/\sigma_p$. This background is eliminated by averaging which enables the averaged JWM to probe phase space at a single point. We note that probing phase space at a single point through averaged measurements is not unique to JWMs. For example, the value of a Wigner function at its origin can be determined by measuring the expectation value of the parity operator $(-1)^{\bm{n}}$, where $\bm{n}$ is the number operator~\cite{royer1977wigner}.

\section{Uncertainty and Predictability}
\label{sec:unc_pred}

In the introduction, we presented weak-measurement as a procedure
that trades away its precision in order to reduce its disturbance. 
However, in the last section we saw that in the weak limit, i.e. $\gamma/\sigma\rightarrow0$, the single-trial
marginal variances go as $\Delta^{2}x\rightarrow 1/(2\sigma^2_p)$ and $\Delta^{2}p\rightarrow \sigma^2_p/2$ (see Eq.~\eqref{eqn:marginals1}).
These are identical to the marginal variances of the final momentum
projector in the JWM. On the surface, it appears that we
have not altered any uncertainties by using weak-measurement.
However, these variances are not the only type of uncertainty that
can appear in a measurement procedure.

Rather than being concerned with the marginal variances of measurement Wigner functions, e.g. $\Delta^{2}x$, weak-measurement trades away
a type of certainty that takes form of $P(x=x'|q')$, the probability
that the system was at $x'$ given outcome $q'$. If the measurement
were strong, the final position of the ancilla would reveal whether
$x=x'$ with certainty. That is, $P(x=x'|q')=0$ or $1$. Conversely,
in the weak limit, $\gamma/\sigma\rightarrow0$, $P(x=x'|q')=1/2$,
equal to a blind guess. This type of certainty has been studied in
which-way and quantum erasure experiments, such as with the double-slit interferometer,
and has been formulated as a measure
called predictability~\cite{greenberger1988simultaneous,englert2000quantitative}.

\begin{figure*}
\centering 
\includegraphics[width=1.0\textwidth]{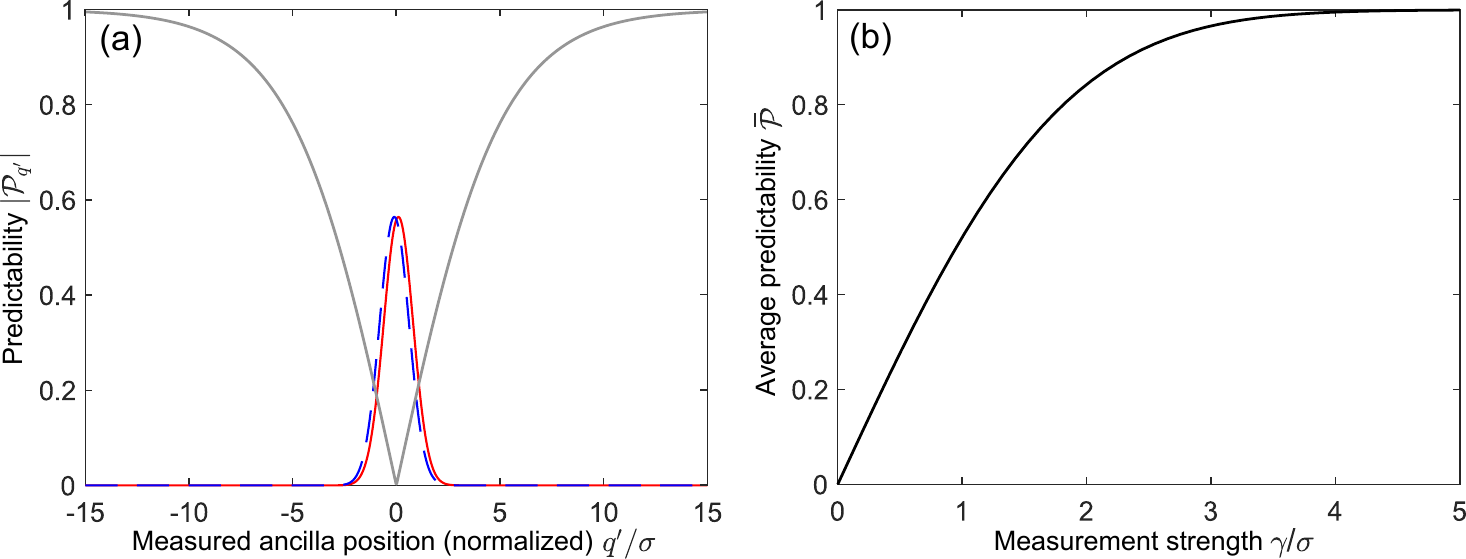}
\caption{Predictability in weak-measurement. (a) The predictability $\left|\mathcal{P}_{q'}\right|$
of whether $x=x'$ or not, given the ancilla is found to have position
$q'$ (grey solid line).
For reference, we also plot the Gaussian ancilla probability distributions given in Eq.~\eqref{eq:prob_pointer},
$P\left(q'|x\protect\neq x'\right)$ and $P\left(q'|x=x'\right)$,
indicated by blue dashed line and red solid line, respectively.
For all the curves, we have set the interaction strength to $\gamma=0.2\sigma$,
where $\sigma$ is the ancilla width. The predictability has the same
form up to a scaling of $q'$ regardless of whether one is in the
weak or strong limit, in that $\left|\mathcal{P}_{q'}\right|=\left|\mathcal{P}_{q'\sigma/\gamma}\right|$.
(b) Finding the expectation of the predictability over all $q'$ gives
the average predictability $\bar{\mathcal{P}}$. For a Gaussian ancilla
probability distribution, $\bar{\mathcal{P}}$ is given by black solid
line. Weak-measurement is defined by $\bar{\mathcal{P}}\ll1$, thereby
ensuring that disturbance to the coherence of the measured system
is minimized. }
\label{fig:predictability}
\end{figure*}

The predictability $\left|\mathcal{P}_{q'}\right|$ is a measure of
how well one can \textit{retrodict} (in our context) whether $x=x'$ given outcome $q'$ (i.e. $P(x=x'|q')$), relative to a blind guess
(i.e. $P(x\neq x')=P(x=x')=1/2$)~\cite{greenberger1988simultaneous}:
\begin{widetext}
\begin{equation}
\begin{split}
\label{eq:full_predictability}
\mathcal{P}_{q'} &=  \frac{P(x=x'|q')-P\left(x=x'\right)}{P\left(x=x'\right)}\\
 &= \frac{P\left(q'|x=x'\right)}{P\left(q'|x=x'\right)P\left(x=x'\right)+P\left(q'|x\neq x'\right)P\left(x\neq x'\right)}-1\\
 &= \frac{P\left(q'|x=x'\right)-P\left(q'|x\neq x'\right)}{P\left(q'|x=x'\right)+P\left(q'|x\neq x'\right)}.
\end{split}
\end{equation}
\end{widetext}
Since we do not know $P(x=x'|q')$, we have used Bayes' law to re-express
it in terms of $P\left(q'|x=x'\right)$ and $P\left(q'|x\neq x'\right)$,
which are given by Eq.~\eqref{eq:prob_pointer}. In Fig.~\ref{fig:predictability}(a), we plot the predictability given by Eq.~\eqref{eq:full_predictability}.
In the strong limit, $P(x=x'|q')=1$ or $0$, so that $\left|\mathcal{P}_{q'}\right|=1$,
as expected. In the weak limit, $\left|\mathcal{P}_{q'}\right|\approx \gamma\left|q'\right|/\sigma^{2} + O\left(\gamma^{3}\right)$
to lowest order in coupling strength $\gamma$. We note that the predictability shares many similarities with other forms of retrodictive certainties studied elsewhere~\cite{appleby1998concept,appleby2000optimal,hofmann2003uncertainty,dressel2014certainty}. These various forms all invoke concepts from retrodictive quantum mechanics~\cite{pegg1999retrodiction,barnett2000bayes,leifer2006quantum,amri2011characterizing,dressel2012contextual,leifer2013towards,dressel2013quantum}. However, distinguishing itself from these other retrodictive certainties, predictability plays a unique role in weak measurement since it explicitly appears in the measurement operator, as we will now show.

Looking back to Eq. \eqref{eqn:approx_proj}, the JWM effectively projects the system
onto a superposition of a momentum eigenstate and, with a relative
amplitude of $\mathcal{P}_{q'}\equiv\gamma q'/\sigma^{2}$, a position
eigenstate. The higher the ability to retrodict whether $x=x'$, the
more the measurement projects onto the corresponding position eigenstate
$\ket{x'^{\mathcal{S}}}$. But predictability has a trade-off. The higher the predictability,
the less coherence is left between the eigenspaces of $\bm{\pi}^{\mathcal{S}}_{x}$
(the $x=x'$ and $x\neq x'$ regions). This trade-off is described
by the wave-particle duality relation $\mathcal{V}_{q'}^{2}+\mathcal{P}_{q'}^{2}\leq1$~\cite{greenberger1988simultaneous, bolduc2014fair}. Here, the coherence has been quantified
by the interference visibility $\mathcal{V}_{q'}$, e.g.
$\mathcal{V}_{q'}=(I_{max}-I_{min})/(I_{max}+I_{min})$ expressed
in terms of the maximum $I_{max}$ and minimum $I_{min}$ intensity of the interference pattern
fringes. If the measured wavefunction $\psi^\mathcal{S}(x)$ had equal amplitudes
for $x=x'$ and $x\neq x'$, e.g. in a double-slit arrangement, they
could initially interfere with perfect visibility, $\mathcal{V}_{q'}$=1.
After the weak-measurement, in the subset of trials for which $q=q'$,
the two amplitudes would interfere with a diminished visibility $\mathcal{V}_{q'}=\sqrt{1-\mathcal{P}_{q'}^{2}}$
\cite{englert2000quantitative,bolduc2014fair}. Consequently, by minimizing the predictability, weak-measurement maximizes the visibility
and thereby maintains coherence in the measured system. That is, it
minimizes disturbance.

It may come as a surprise that, even in the weak limit, the predictability
$\left|\mathcal{P}_{q'}\right|$ can become significant if $|q'|>\sigma^{2}/\gamma$.
For these outlier ancilla position outcomes, we can retrodict with certainty
whether the system was at $x'$ or not. To understand this, consider
if the ancilla's position is found to be zero, i.e. $q'=0$. Because
the ancilla's Gaussian probability density $|\phi^{\mathcal{A}}(q)|^{2}$ is relatively
constant near its center at $q=0$, in the weak limit, $\gamma/\sigma\ll1$,
the probability for this outcome is the same regardless of whether
$x=x'$ or $x\neq x'$ . Consequently, little information is acquired
about the position of the system in that trial. In contrast, consider
if one measures the ancilla's position to be many standard deviations
away from $q'=0$ in the positive direction, i.e. $q'\gg\sigma$. Due
to the exponential decay of $|\phi^{\mathcal{A}}(q)|^{2}$, this outcome
occurs with a higher probability in the case where $x=x'$ (the ancilla is shifted) than the case $x\neq x'$ (the ancilla is not shifted). Similarly, the reverse is true when $q'\ll-\sigma$.
Thus, for these outlier outcomes of $q'$ one can be relatively certain
of whether $x=x'$ or not. In this way, the retrodictive certainty
of the weak-measurement varies from trial-to-trial.


While the ancilla outcomes with large $q'$ give high predictability,
they occur rarely since they lie in the tails of the Gaussian $|\phi^{\mathcal{A}}(q)|^2$. The majority
of the trials give low predictability. To incorporate this effect
we consider the predictability averaged over all the final ancilla
positions (following \cite{bolduc2014fair}),
\begin{widetext}
\begin{equation}
\begin{aligned}
\label{eq:avg_predictability}
\bar{\mathcal{P}}= & \intop dq'\left|\mathcal{P}_{q'}\right|P(q')\\
= & \intop dq'\left|\mathcal{P}_{q'}\right|\left(P\left(q'|x=x'\right)P\left(x=x'\right)+P\left(q'|x\neq x'\right)P\left(x\neq x'\right)\right)\\
= & \frac{1}{2}\intop dq'\left|\mathcal{P}_{q'}\right|\left(P\left(q'|x=x'\right)+P\left(q'|x\neq x'\right)\right),
\end{aligned}
\end{equation}
\end{widetext}
where we have used Bayes' law again and $P(x\neq x')=P(x=x')=1/2$.
We plot the average predictability $\bar{\mathcal{P}}$ in Fig.~\ref{fig:predictability}(b). Similarly, the average
visibility $\bar{\mathcal{V}}$ will be limited to $\bar{\mathcal{V}}\leq\sqrt{1-\bar{\mathcal{P}}^{2}}$
\cite{englert2000quantitative,bolduc2014fair}. In the strong limit,
$\bar{\mathcal{P}}\rightarrow1$. In the weak limit, $\bar{\mathcal{P}}=\gamma /(\sigma \sqrt{\pi}) +\mathcal{O}(\gamma^{3}).$
Consequently, for the average induced disturbance of the weak-measurement
to be small, one must have $\gamma/\sigma\ll1$, which is the standard
weakness condition.

Our result can be related the question debated in Refs.~\cite{Scully1991quantum,Storey1994path,Wiseman1995uncertainty} of whether momentum disturbance is needed to erase interference fringes in a double-slit interferometer. The relation $\bar{\mathcal{V}}\leq\sqrt{1-\bar{\mathcal{P}}^{2}}$ explicitly reveals the trade-off between which-way information and fringe visibility. In particular, the ability to retrodict which slit a photon went through, quantified by $\bar{\mathcal{P}}$, comes at the cost of reduced interference fringe visibility $\bar{\mathcal{V}}$, as enforced by complementarity. We also find that the predictability depends explicitly on the which-way measurement strength, i.e. $\bar{\mathcal{P}}=\gamma /(\sigma \sqrt{\pi})$, and hence on its disturbance. Thus, as in Refs.~\cite{Wiseman1995uncertainty,mir2007double}, we conclude that both measurement disturbance and complementarity play a role in the trade-off between which-way information and fringe visibility.

\section{Conclusions}
The Heisenberg uncertainty principle is often interpreted as a trade-off between the precision and disturbance of a measurement. Gentle or ``weak-measurements" can be used to minimize such measurement-induced disturbance. Here we studied a joint weak-measurement (JWM) consisting of a weak position measurement followed by a regular momentum measurement. One intuitively expects that the weak-measurement's reduction in disturbance comes at the cost of a reduction in its precision. While this intuition is correct for a single JWM trial, we showed that averaging over many trials compensates for the loss in certainty of the weak-measurement. This enables the average outcome of a JWM to probe phase space with a precision exceeding the uncertainty principle limit. The weak position measurement does not trade away the usual notion of certainty, i.e. the standard deviation $\Delta x$ that appears in the uncertainty principle. Rather, it trades away the certainty with which one can retrodict the outcome of the measurement.

JWMs have already found numerous applications in quantum physics. For instance, they have been used to study foundational topics such as testing error-disturbance relations~\cite{rozema2012violation,ringbauer2014joint}, resolving quantum paradoxes~\cite{lundeen2009experimental,yokota2009direct}, and reconstructing Bohmian trajectories~\cite{kocsis2011observing,mahler2016experimental}. Moreover, JWMs have been used to directly determine quantum states~\cite{Lundeen2011direct,bamber2014observing,thekkadath2016direct}, a technique especially useful to efficiently characterize high-dimensional systems~\cite{malik2014direct,shi2015scan}. Despite the fact that JWMs are being increasingly used in quantum physics experiments, there were questions regarding their compatibility with fundamental concepts such as the uncertainty principle and complementarity. Our results answer these questions and provide an intuitive understanding of the mechanism behind JWMs.

\begin{acknowledgements}
We thank A.M. Steinberg and J.H. Shapiro for the initial discussions that prompted this work. We also thank J. Sperling for their insightful comments on the manuscript. This work was supported by the Canada Research Chairs (CRC) Program, the Canada First Research Excellence Fund (CFREF), and the Natural Sciences and Engineering Research Council (NSERC). G.S.T acknowledges support from the Oxford Basil Reeve Graduate Scholarship.
\end{acknowledgements}

\newpage
\onecolumngrid

\appendix
\section{Projector expansion of joint weak-measurement}
\label{app:A} 
Here we show that the JWM operator $\bm{M}_{q',x',p'}^{\mathcal{S}}$ can be expressed
as a simultaneous projection onto position and momentum eigenstates.
The operator is defined as $\bm{M}_{q',x',p'}^{\mathcal{S}}=\bra{\phi^{\mathcal{A}}}\bm{U}_{x'}^{\dagger}\bm{\pi}_{q'}^{\mathcal{A}}\bm{\pi}_{p'}^{\mathcal{S}}\bm{U}_{x'}\ket{\phi^{\mathcal{A}}}$,
which can be written as an unnormalized projector $\bm{M}_{q',x',p'}^{\mathcal{S}}=\ket{\theta}\bra{\theta}$
where 
\begin{equation}
\ket{\theta}=\bra{\phi^{\mathcal{A}}}\bm{U}_{x'}^{\dagger}\ket{q'^{\mathcal{A}}}\ket{p'^{\mathcal{S}}}.
\label{eqn:theta}
\end{equation}
The unitary can be expanded $\bm{U}_{x'}^{\dagger}=\exp{(-\gamma\bm{\pi}_{x'}^{\mathcal{S}}\otimes\partial_{q'}^{\mathcal{A}})}=\mathbbm{1}^{\mathcal{S}}\otimes\mathbbm{1}^{\mathcal{A}}+\bm{\pi}_{x'}^{\mathcal{S}}\sum_{n=1}^{\infty}(-\gamma\partial_{q'}^{\mathcal{A}})^{n}/n!$.
Inserting this expression into Eq.~\eqref{eqn:theta}, we find: 
\begin{equation}
\begin{split}
\ket{\theta} & =\phi^{\mathcal{A}}(q')\ket{p'^{\mathcal{S}}}+e^{ip'x'}\left(\sum_{n=1}^{\infty}\frac{1}{n!}(-\gamma\partial_{q'}^{\mathcal{A}})^{n}\phi^{\mathcal{A}}(q')\right)\ket{x'^{\mathcal{S}}}\\
 & =\phi^{\mathcal{A}}(q')\left[ \ket{p'^{\mathcal{S}}} + e^{ip'x'}\left(\sum_{n=1}^{\infty}\frac{1}{n!}\left(\frac{\gamma}{\sigma}\right)^{n}He_{n}\left(\frac{q'}{\sigma}\right)\right)\ket{x'^{\mathcal{S}}} \right] \\
 & \equiv \phi^{\mathcal{A}}(q') \ket{x'^{\mathcal{S}},p'^{\mathcal{S}}}_{q'}
\end{split}
\end{equation}
where we used the fact that $\braket{x'^{\mathcal{S}}|p'^{\mathcal{S}}}=e^{ip'x'}$
and $\partial_{q'}^{n}(e^{-q'^{2}/2\sigma^{2}})=(\frac{-1}{\sigma})^{n}He_{n}\left(\frac{q'}{\sigma}\right)e^{-q'^{2}/2\sigma^{2}}$
where $He_{n}$ are the so-called probabilists' Hermite polynomials.
In the weak measurement limit $\gamma/\sigma\ll1$, we consider only
the $n=1$ term in the sum in which $He_{1}(q')=q'$, thus yielding:
\begin{equation}
\ket{x'^{\mathcal{S}},p'^{\mathcal{S}}}_{q'}\approx \ket{p'^{\mathcal{S}}}+\left(\frac{\gamma q'}{\sigma^{2}}\right)e^{ip'x'}\ket{x'^{\mathcal{S}}}
\label{eqn:approx_ket}
\end{equation}
which is Eq.~\eqref{eqn:approx_proj} in the main text.

\section{Generalizing the joint weak-measurement}
\label{app:B}
Here we generalize the JWM to take into account projectors having a finite width. That is, the JWM sequence now consists of a weak-measurement of $\bm{\Pi}^{\mathcal{S}}_{x'} = \ket{\chi^{\mathcal{S}}(x')}\bra{\chi^{\mathcal{S}}(x')}$ and a regular measurement of $\bm{\Pi}^{\mathcal{S}}_{p'} = \ket{\Gamma^{\mathcal{S}}(p')}\bra{\Gamma^{\mathcal{S}}(p')}$ where $\chi^{\mathcal{S}}(x)$ and $\Gamma^{\mathcal{S}}(p)$ are finite-width distributions. Using the same reasoning as before, we write the JWM operator $\bm{M}_{q',x',p'}^{\mathcal{S}}=\bra{\phi^{\mathcal{A}}}\bm{U}_{x'}^{\dagger}\bm{\pi}_{q'}^{\mathcal{A}}\bm{\pi}_{p'}^{\mathcal{S}}\bm{U}_{x'}\ket{\phi^{\mathcal{A}}}$ as a projector $\ket{\theta}\bra{\theta}$ where $\ket{\theta} = \braket{\phi^\mathcal{A}|\bm{U}_{x'}^\dagger|q'^{\mathcal{A}}}\ket{\Gamma^{\mathcal{S}}(p')}$ and $\bm{U}_{x'}^\dagger = \exp{\left(-\gamma \ket{\chi^{\mathcal{S}}(x')}\bra{\chi^{\mathcal{S}}(x')} \otimes \partial^\mathcal{A}_{q'}\right)}$. In the weak measurement limit $\gamma / \sigma \ll 1$, the unitary $\bm{U}_{x'}^\dagger$ can be expanded to first order which leads to:
\begin{equation}
\ket{\Delta^{\mathcal{S}}_{q'}(x',p')} \approx \ket{\Gamma^{\mathcal{S}}(p')} + \mathcal{P}_{q'} \braket{\chi^{\mathcal{S}}(x')|\Gamma^{\mathcal{S}}(p')} \ket{\chi^{\mathcal{S}}(x')}.
\label{eqn:generalized_DM_state}
\end{equation}
such that $\bm{M}_{q',x',p'}^{\mathcal{S}}=|\phi^\mathcal{A}(q')|^2 \ket{\Delta^{\mathcal{S}}_{q'}(x',p')}\bra{\Delta^{\mathcal{S}}_{q'}(x',p')}$. 

We now assume that $\chi^{\mathcal{S}}(x) = e^{-x^2/2\sigma_x^2}/(\pi\sigma_x^2)^{1/4}$ and $\Gamma^{\mathcal{S}}(p) = e^{-p^2/2\sigma_p^2}/(\pi\sigma_p^2)^{1/4}$ are Gaussians of widths $\sigma_x$ and $\sigma_p$, respectively. Then, $\braket{\chi^{\mathcal{S}}(x')|\Gamma^{\mathcal{S}}(p')} = \int_{-\infty}^{\infty} dx \chi^{\mathcal{S}}(x-x')\tilde{\Gamma}^{\mathcal{S}}(x)e^{ip'x}$ where $\tilde{\Gamma}^{\mathcal{S}}(x)$ is the Fourier transform of $\Gamma^{\mathcal{S}}(p)$. When $\sigma_x\sigma_p \ll 1$, this quantity can be approximated as $\braket{\chi(x')^{\mathcal{S}}|\Gamma(p')^{\mathcal{S}}} \approx e^{ip'x'}\sqrt{2\sigma_x\sigma_p}$. In this limit:
\begin{equation}
\ket{\Delta^{\mathcal{S}}_{q'}(x',p')} \approx \ket{\Gamma^{\mathcal{S}}(p')} + \mathcal{P}_{q'}\sqrt{2\sigma_x\sigma_p}e^{ip'x'}\ket{\chi^{\mathcal{S}}(x')}.
\label{eqn:generalized_DM_state2}
\end{equation}
which is Eq.~\eqref{eqn:approx_proj_general} in the main text.

\section{Wigner function of the joint weak-measurement}
\label{app:C}
Here we derive the Wigner function $W_{\bm{M}}(x,p)$ of the generalized JWM by computing the inverse Weyl transform of $\bm{M}_{q',x',p'}^{\mathcal{S}} = |\phi^\mathcal{A}(q')|^2 \ket{\Delta^{\mathcal{S}}_{q'}(x',p')}\bra{\Delta^{\mathcal{S}}_{q'}(x',p')}$ where $\ket{\Delta^{\mathcal{S}}_{q'}(x',p')}$ is given in Eq.~\eqref{eqn:generalized_DM_state2}. We consider the case  $(x',p')=(0,0)$ such that $\ket{\Delta^{\mathcal{S}}_{q'}(0,0)} = \ket{\Gamma^{\mathcal{S}}(0)} + \mathcal{P}_{q'}\sqrt{2\sigma_x\sigma_p}\ket{\chi^{\mathcal{S}}(0)}$. The Wigner function $W_{\bm{M}}(x,p)$ consists of three terms:
\begin{equation}
W_{\bm{M}}(x,p) = |\phi^\mathcal{A}(q')|^2 \left (  W_{\Gamma}(x,p) + \mathcal{P}^2_{q'} 2 \sigma_x \sigma_p W_{\chi}(x,p) + \left( \mathcal{P}_{q'}\sqrt{2\sigma_x\sigma_p}W_{\Gamma\chi}(x,p) + c.c. \right) \right ).
\end{equation}
The first term, $W_{\Gamma}(x,p)$, is given by:
\begin{equation}
\begin{split}
W_{\Gamma}(x,p) &= \frac{1}{\pi} \int_{-\infty}^{\infty}dy \tilde{\Gamma}^{\mathcal{S}}(x+y)\tilde{\Gamma}^{\mathcal{S}}(x-y) e^{-i2py} \\
&= \left(\frac{\sigma_p^2}{\pi^3}\right) ^{1/2} \int_{-\infty}^{\infty}dy e^{-(x+y)^2\sigma_p^2/2}e^{-(x-y)^2\sigma_p^2/2}e^{-i2py} \\
&= e^{-(x^2\sigma_p^4 + p^2)/\sigma_p^2} / \pi
\end{split}
\end{equation}
which is a squeezed vacuum state with $\Delta x = 1/\sqrt{2}\sigma_p$ and $\Delta p = \sigma_p /\sqrt{2}$~\cite{leonhardt1997measuring}. The second term, $W_{\chi}(x,p)$, is given by:
\begin{equation}
\begin{split}
W_{\chi}(x,p) &= \frac{1}{\pi} \int_{-\infty}^{\infty}dy \chi^{\mathcal{S}}(x+y)\chi^{\mathcal{S}}(x-y) e^{-i2py} \\
&= \left( \frac{1}{\pi^3\sigma_x^2} \right)^{1/2} \int_{-\infty}^{\infty}dy e^{-(x+y)^2/2\sigma^2_x} e^{-(x-y)^2/2\sigma^2_x}e^{-i2py} \\
&= e^{-(x^2 + p^2\sigma_x^4)/\sigma_x^2} / \pi
\end{split}
\end{equation}
which is a squeezed vacuum state with $\Delta x = \sigma_x/\sqrt{2}$ and $\Delta p = 1/\sqrt{2}\sigma_x$. Finally, the third term, $W_{\Gamma\chi}(x,p)$, is given by:
\begin{equation}
\begin{split}
W_{\Gamma\chi}(x,p) &= \frac{1}{\pi} \int_{-\infty}^{\infty}dy \chi^{\mathcal{S}}(x+y)\tilde{\Gamma}^{\mathcal{S}}(x-y) e^{-i2py} \\
&= \left( \frac{\sigma^2_p}{\pi^6\sigma_x^2} \right)^{1/4} \int_{-\infty}^{\infty}dy e^{-(x+y)^2/2\sigma^2_x} e^{-(x-y)^2\sigma_p^2/2} e^{-i2py} \\
&\approx \sqrt{2\sigma_x\sigma_p} e^{-2(x^2\sigma_x^2 + p^2\sigma_p^2)} e^{2ipx} / \pi + \mathcal{O}(\sigma_x^2 \sigma_p^2)
\end{split}
\end{equation}
where we ignored $\mathcal{O}(\sigma_x^2 \sigma_p^2)$ terms since $\sigma_x\sigma_p \ll 1$. Combining these results, we obtain the JWM Wigner function:
\begin{equation}
\label{eqn:general_wig}
W_{\bm{M}}(x,p) = \left|\phi^\mathcal{A}(q')\right|^2\left( e^{-(x^2\sigma_p^4 + p^2)/\sigma_p^2} + 2\mathcal{P}^2_{q'}\sigma_x\sigma_p e^{-(x^2 + p^2\sigma_x^4)/\sigma_x^2} + 4\mathcal{P}_{q'}\sigma_x\sigma_p e^{-2x^2\sigma_p^2 - 2p^2\sigma_x^2}\cos{(2xp)} \right) / \pi,
\end{equation}
which is plotted in Fig.~\ref{fig:3dWigner}.

\section{Marginals of the joint weak-measurement}
\label{app:D}
Here we derive the marginals of the JWM Wigner function $W_{\bm{M}}(x,p)$, that is $P(x) = \left|\braket{x|\Delta^{\mathcal{S}}_{q'}(x',p')}\right|^2$ and $P(p) = \left|\braket{p|\Delta^{\mathcal{S}}_{q'}(x',p')}\right|^2$ (note that these marginals are divided by $\left|\phi^\mathcal{A}(q')\right|^2$ so that they are normalized). We compute these directly from Eq.~\eqref{eqn:generalized_DM_state2}:
\begin{equation}
\label{eqn:general_marginals}
\begin{split}
P(x) &= \left|\tilde{\Gamma}^{\mathcal{S}}(x)\right|^2 + 2\sigma_x\sigma_p\mathcal{P}^2_{q'}\left|\chi^{\mathcal{S}}(x-x')\right|^2 + 2\mathcal{P}_{q'}\sqrt{2\sigma_x\sigma_p}\chi^{\mathcal{S}}(x-x')\tilde{\Gamma}^{\mathcal{S}}(x)\cos{(p'(x-x'))} \\
P(p) &=\left|\Gamma^{\mathcal{S}}(p-p')\right|^2 + 2\sigma_x\sigma_p\mathcal{P}^2_{q'}\left|\tilde{\chi}^{\mathcal{S}}(p)\right|^2 + 2\mathcal{P}_{q'}\sqrt{2\sigma_x\sigma_p}\Gamma^{\mathcal{S}}(p-p')\tilde{\chi}^{\mathcal{S}}(p)\cos{(x'(p-p'))}
\end{split}
\end{equation}
where $\tilde{\chi}^{\mathcal{S}}(p)$ is the Fourier transform of $\chi^{\mathcal{S}}(x)$. In general, the variances of $P(x)$ and $P(p)$ will depend on where the system is being probed, i.e. $(x',p')$. However, in practice this change is negligible since the JWM experimental apparatus typically operates in the regime $x'_{max} \ll 1/\sigma_p$ and $p'_{max}\ll 1/\sigma_x$, where $x'_{max}$ and $p'_{max}$ are the respective spatial and momentum extent of the phase space area probed, i.e. $x'\in[-x'_{max},x'_{max}]$ and $p'\in[-p'_{max},p'_{max}]$~\cite{Lundeen2011direct}. Satisfying these two conditions ensures sufficient measurement precision relative to the characteristic size of the system. Moreover, in this regime the measurement probe location $(x',p')$ simply shifts the center position of the marginals without changing their shape. Thus, the variances can be determined directly from the second moment of $P(x)$ and $P(p)$ for $(x',p')=(0,0)$:
\begin{equation}
\begin{split}
\Delta^2 x &= \int^{\infty}_{-\infty} dx  x^2 \left( \left|\tilde{\Gamma}^{\mathcal{S}}(x)\right|^2 + 2\sigma_x\sigma_p\mathcal{P}^2_{q'}\left|\chi^{\mathcal{S}}(x)\right|^2 + 2\mathcal{P}_{q'}\sqrt{2\sigma_x\sigma_p}\chi^{\mathcal{S}}(x)\tilde{\Gamma}^{\mathcal{S}}(x) \right)  \approx 1/(2\sigma^2_p) + \mathcal{P}_{q'}^2\sigma_x^3\sigma_p + 4\mathcal{P}_{q'}\sigma_x^3\sigma_p \\
\Delta^2 p  &= \int^{\infty}_{-\infty} dp p^2 \left(\left|\Gamma^{\mathcal{S}}(p)\right|^2 + 2\sigma_x\sigma_p\mathcal{P}^2_{q'}\left|\tilde{\chi}^{\mathcal{S}}(p)\right|^2 + 2\mathcal{P}_{q'}\sqrt{2\sigma_x\sigma_p}\Gamma^{\mathcal{S}}(p)\tilde{\chi}^{\mathcal{S}}(p) \right) \approx \sigma^2_p/2 + \mathcal{P}_{q'}^2\sigma_p/\sigma_x + 4\mathcal{P}_{q'}\sigma_p^3\sigma_x,
\end{split}
\end{equation}
where we used the approximation $1/(\sigma^2_x\sigma^2_p + 1) \approx 1$.

\bibliographystyle{apsrev4-1}
\bibliography{refs}

\end{document}